\documentclass[conference,10pt]{IEEEtran}
\IEEEoverridecommandlockouts
\usepackage{cite}
\usepackage{amsmath,amssymb,amsfonts}
\usepackage{graphicx}
\usepackage{textcomp}
\usepackage{caption}
\usepackage{subcaption}
\usepackage{xcolor}
\usepackage{hyperref}
\def\BibTeX{{\rm B\kern-.05em{\sc i\kern-.025em b}\kern-.08em
    T\kern-.1667em\lower.7ex\hbox{E}\kern-.125emX}}
\usepackage[norelsize, linesnumbered, ruled, lined, boxed, commentsnumbered, noend]{algorithm2e}
\usepackage{setspace}
\usepackage{fancyhdr}
\usepackage{comment}
\usepackage[most]{tcolorbox}
\usepackage{xcolor}
\usepackage{soul}
\usepackage{bm}

\definecolor{appgreen}{HTML}{60A917} 
\definecolor{appdarkred}{HTML}{FF0000} 
\definecolor{applightred}{HTML}{F8CECC} 

\definecolor{machineblue}{HTML}{DAE8FC} 
\definecolor{machineyellow}{HTML}{FFF2CC} 
\definecolor{machineorange}{HTML}{FFE6CC} 
\definecolor{machineviolet}{HTML}{E1D5E7} 

\newtcbox{\highlightappgreen}[1][appgreen]{
  on line,
  arc=0pt,
  outer arc=0pt,
  colback=#1,
  colframe=black,
  boxrule=0.5mm,
  boxsep=0pt,
  left=1mm,
  right=1mm,
  top=1mm,
  bottom=1mm
}
\newtcbox{\highlightappdarkred}[1][appdarkred]{
  on line,
  arc=0pt,
  outer arc=0pt,
  colback=#1,
  colframe=black,
  boxrule=0.5mm,
  boxsep=0pt,
  left=1mm,
  right=1mm,
  top=1mm,
  bottom=1mm
}
\newtcbox{\highlightapplightred}[1][applightred]{
  on line,
  arc=0pt,
  outer arc=0pt,
  colback=#1,
  colframe=black,
  boxrule=0.5mm,
  boxsep=0pt,
  left=1mm,
  right=1mm,
  top=1mm,
  bottom=1mm
}

\newtcbox{\highlightmachineblue}[1][machineblue]{
  on line,
  arc=0pt,
  outer arc=0pt,
  colback=#1,
  colframe=black,
  boxrule=0.5mm,
  boxsep=0pt,
  left=1mm,
  right=1mm,
  top=1mm,
  bottom=1mm
}

\newtcbox{\highlightmachineyellow}[1][machineyellow]{
  on line,
  arc=0pt,
  outer arc=0pt,
  colback=#1,
  colframe=black,
  boxrule=0.5mm,
  boxsep=0pt,
  left=1mm,
  right=1mm,
  top=1mm,
  bottom=1mm
}

\newtcbox{\highlightmachineorange}[1][machineorange]{
  on line,
  arc=0pt,
  outer arc=0pt,
  colback=#1,
  colframe=black,
  boxrule=0.5mm,
  boxsep=0pt,
  left=1mm,
  right=1mm,
  top=1mm,
  bottom=1mm
}
\newtcbox{\highlightmachineviolet}[1][machineviolet]{
  on line,
  arc=0pt,
  outer arc=0pt,
  colback=#1,
  colframe=black,
  boxrule=0.5mm,
  boxsep=0pt,
  left=1mm,
  right=1mm,
  top=1mm,
  bottom=1mm
}
\begin{document}
\pagestyle{fancy}

\author{\IEEEauthorblockN{Suvarthi Sarkar, Nandini Sharma, Akshat Mittal, Aryabartta Sahu \textit{IEEE Senior Member}}
\IEEEauthorblockA{\textit{Dept. of CSE,  IIT Guwahati, Assam, India.} Emails:\{s.sarkar, s.nandini, m.akshat, asahu\}@iitg.ac.in }
}

\title{Power Aware Container Placement in Cloud Computing with Affinity and Cubic Power Model}

\renewcommand{\algorithmautorefname}{Algorithm}
\maketitle

\begin{abstract}
Modern data centres are increasingly adopting containers to enhance power and performance efficiency. These data centres consist of multiple heterogeneous machines, each equipped with varying amounts of resources such as CPU, I/O, memory, and network bandwidth. Data centers rent their resources to applications, which demand different amounts of resources and execute on machines for extended durations if the machines provide the demanded resources to the applications. Certain applications run efficiently on specific machines, referred to as system affinity between applications and machines. In contrast, others are incompatible with specific machines, referred to as anti-affinity between applications and machines. We consider that there are multiple applications, and data centers need to execute as many applications as possible. Data centers incur electricity based on CPU usage due to the execution of applications, with the cost being proportional to the cube of the total CPU usage. It is a challenging problem to place applications on the machines they have an affinity for while keeping the electricity cost in check. 
Our work addresses the placement problem of matching applications to machines to minimize overall electricity costs while maximizing the number of affinity pairs of machines and applications. We propose three solution approaches: (a) Power-Aware Placement (PAP): applications are placed on machines where power usage is minimized, (b) Affinity-Aware Placement (AAP): applications are placed on machines where affinity is maximized, (c) Combined Power-Affinity Placement (CPAAP): this approach integrates the benefits of both PAP and AAP. Our proposed approach improves the affinity satisfaction ratio by up to 4\% while reducing the total system cost by up to 26\% and improving the affinity payoff ratio by up to 37\% compared to state-of-the-art approaches for real-life datasets.

\end{abstract}


\setstretch{0.9}

\section{Introduction}
Almost 60\% of the operational cost of a data center (DC) is spent on electricity consumption \cite{yi2021, CHEUNG18}. Meanwhile, the electricity demand continues to rise. The cloud computing industry aims to maximize resource utilization while minimizing energy use. Placing application containers on the cloud must be energy-efficient and address issues that arise when multiple applications share the same cloud space, interfering and competing for resources. Power consumption at cloud DCs should be minimized while maintaining smooth operations.

Conventional DCs typically provide users with various virtual machines (VMs) that are packaged with cloud resources \cite{global_cost_aware}. These DCs have poor resource utilisation and excessive power consumption. Thus, it is imperative that cloud service providers encourage the transition from older DCs to more advanced DCs with increased processing and energy efficiency. Container cloud platforms have gained popularity as a low-weight substitute for VMs in recent years due to their many benefits, including elastic scaling, high resource utilisation, and quick startup times \cite{kubernetes}. This efficiency allows for higher density and better resource utilization, enabling more applications to run on the same hardware.

\begin{figure}[tb!]
    \centering
    \includegraphics[scale=1]{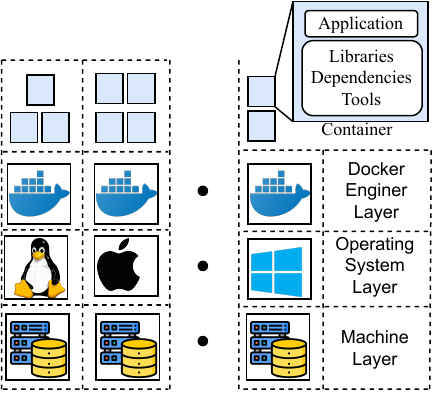}
    \caption{\footnotesize Relationship between containers and applications}
    \label{fig:container}
\end{figure}

The advent of container technology has significantly generalized the deployment of large internet-based applications, which often require numerous instances distributed across multiple servers in a cluster \cite{global_cost_aware}. Containers offer a consistent and isolated environment for running applications, ensuring portability and efficient resource utilization. On the other hand, applications are software programs that provide specific functionalities to users. The container requirements are closely tied to the characteristics and needs of the applications they host. Each application typically runs within a single container, while each machine in the cluster can host multiple containers, potentially running applications of the same or different types. Each machine runs an operating system on which the Docker Engine is installed. The Docker Engine manages and hosts all the containers \cite{docker}. Each container includes a distinct application along with all necessary libraries and dependencies. \autoref{fig:container} shows the container and application relationship. In this work, we address the problem of placing containers on machines. Since applications are containerized before placement, we use the term ``placement of applications on machines" rather than ``placement of containers on machines."

Given the heterogeneity of machines in terms of available resources and the diverse requirements of different containers, it is crucial to place containers on machines to balance resource utilization strategically. Furthermore, constraints such as affinity and anti-affinity make the placement problem more complex, where certain groups of containers are best placed together on specific machines. In contrast, others must be kept separate due to various constraints.

Affinity means the application must be deployed on machines with specific kernel versions or in specific resource pools due to performance or compliance \cite{moreno2018}. For example, network-intensive applications should only be installed on hosts with high network bandwidth, and graphics-intensive applications should be run on machines with good graphics processing units. Conversely, anti-affinity refers to the inability of an application to be installed on certain computers that lack the hardware and software requirements specified by the application. For instance, hosting in or out of a specific location might be restricted due to legal requirements \cite{monitoring_audit_logging}, the necessity to ensure certain services are closer to the user's location, lack of security permissions, or incompatibility between the machine and the software. \cite{espling2016}. Thus, affinity and anti-affinity play a crucial role in lowering power usage and increasing the operating efficiency of DC.

In this work, we propose a combine power consumption and affinity aware placement (CPAAP) approach. The goal is to minimize the total cost of the cluster, which involves power consumption and affinity satisfaction of applications. It is essential to understand the challenges faced during placement. We consider this placement is done by the container mapper, which simply decides which application is placed on which machine. Due to the anti-affinity relationship of applications and machines, container mapper face the challenge of allocating resources to an application request, even if they are available in the cluster. At the same time, the container mapper also needs to allocate power-efficient machines to balance the total power consumption of DC. Since containers remain on a machine for a long time, it is essential to use an effective placement approach to maximize the profit of DC.

The main contributions of the work are as follows:

\begin{itemize}
    \item In this work, we tackle the challenge of allocating applications to machines by considering multiple resources, affinity and anti-affinity factors, and a cubic power model. While existing literature offers solutions addressing either affinity and anti-affinity factors \cite{global_cost_aware}, or multiple resources \cite{group-scheduling, Bian19}, or the cubic power model \cite{Kaur17}, the integration of all three factors has not been previously explored. This work pioneers the combination of these elements for application allocation.

    \item We proposed a method to derive the affinity between a machine and an application by considering the application's resource demand and the resources available to be placed in the machine for the application.

    \item We provided a proper problem formulation and simulated the proposed approach on synthetic and real-life application traces. Our approach outperforms the existing state-of-the-art solutions and reduces system cost by 26\%.

\end{itemize}

\section{Related Works}
VM placement is a primary focus of cloud computing research, aiming to optimize resource allocation, enhance performance, and reduce energy consumption. Alboaneen \emph{et al.} \cite{meta_heuristic_joint_task_scheduling}proposed a meta-heuristic approach to maximize data centre workload placement and VM placement. The goal of the technique is to assign tasks to the VM with the lowest execution cost under the deadline constraint. Lin \emph{et al.} \cite{Lin22} proposed a strategy that dynamically consolidates VMs based on real-time resource utilization metrics, optimizing server cluster performance and reducing energy consumption.

A more sophisticated form of virtualization is containerization. It is simple to monitor and administer because it is lighter and more transparent than VMs. Google has an open-source container orchestration system called Kubernetes \cite{kubernetes} that facilitates containerized application administration, large-scale scaling and automated deployment. The application's default resource configuration technique modifies the number of containers it runs in accordance with CPU utilisation.

Saving power and protecting the environment are always critical concerns in placement studies. Menouer \emph{et al.} \cite{power_container_ml} proposed a machine learning-based solution for power-efficient container placement for cloud computing environments. Zhou \emph{et al.} \cite{Zhou19}, Xu \emph{et al.} \cite{Xu19} proposed load balancing techniques to lower the power consumption of the DC.

Node affinity is an interesting feature that Kubernetes \cite{docker-kubernetes-scheduling} added to handle more complicated placement circumstances. Hu \emph{et al.} \cite{concurrent_container_heterogeneous} proposed to put containers on an affinity machine and containers with mutual affinity on the same machine to reduce network transmission time. They employed an easily adjustable flow network to show these affinity links and determine the optimal routing scheme based on allocation costs. 
Tan \emph{et al.} \cite{Tan22} introduced an innovative availability-aware container scheduler that prioritizes cloud environments' service availability and fault tolerance. By integrating redundancy, dynamic adaptation, and efficient resource utilization, the scheduler enhances the reliability of application services, ensuring minimal downtime and robust performance.

% \subsection{Existing Work}
In Long \emph{et al.} \cite{global_cost_aware}, they proposed a Global Cost-Aware Scheduling approach (GCCS) to address the container instance allocation problem in heterogeneous server clusters. By formulating the issue as an integer linear programming (ILP), the authors introduce a heuristic search approach to optimize the allocation of containers to machines, considering affinity and anti-affinity requirements. Notably, the study employs a Bayesian optimizer to automate the determination of the cost coefficient, a crucial parameter in the placement approach. Results show that GCCS significantly reduces total power consumption and maintains a high affinity satisfaction ratio.

% \subsection{Conclusion}
While the existing works have considered power optimization along with affinity/anti-affinity satisfaction, they have taken the affinity/anti-affinity to be binary \cite{global_cost_aware}. The power model studied in their work is linear. Also, only the computational power of machines is considered by taking only the number of CPU cores in machines and CPU cores required by applications. Our work differs from previous works as we explore non-binary affinity values and consider additional resources such as network, memory and I/O of machines and applications. We use the more realistic cubic power consumption model \cite{cubic_power_model}. We then explore the problem of maximizing affinity satisfaction, which leads to improved performance and minimizes the cost due to power consumption.

\section{System Model}

\begin{figure}
    \centering
    \includegraphics[scale=.83]{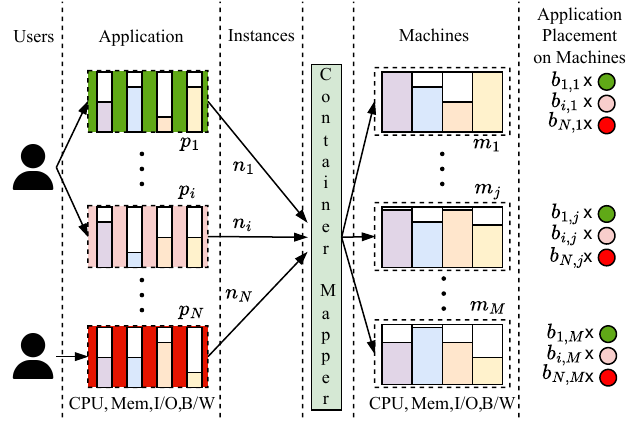}
    \caption{\footnotesize Container Instance Placement Model. The applications are marked in different colours: $p_1$ with \highlightappgreen{\textcolor{appgreen}{H}}, $p_i$ with \highlightapplightred{\textcolor{applightred}{H}} and $p_N$ with \highlightappdarkred{\textcolor{appdarkred}{H}}. Each applications are filled with a cylinder indicating the resource availability and requirements: \highlightmachineviolet{\textcolor{machineviolet}{H}} indicates CPU, \highlightmachineblue{\textcolor{machineblue}{H}} indicates memory, \highlightmachineorange{\textcolor{machineorange}{H}} indicates I/O, \highlightmachineyellow{\textcolor{machineyellow}{H}} indicates B/W.}
    \label{fig:system-model}
\end{figure}

In this section, we describe the system model of the container instance placement problem, which is shown in \autoref{fig:system-model}. The machine nodes of the cloud environment are defined as the set \(\boldsymbol{\Delta} = \{ m_1, m_2, \cdots, m_j, \cdots, m_M \}\), where M is the number of machine nodes, and $m_j$ represents the $j^{th}$ machine. We use the variable $j$ to identify a particular machine uniquely. The resource capacity of the machines is given by the set \(\emph{\textbf{V}} = \{ v_1, v_2,\cdots, v_j,\cdots, v_M \}\), where \(v_j\) represents the resource of machine $m_j$, $v_j$ is a tuple. It is represented by \(\left(cpu_j^{cap}, io_j^{cap}, nw_j^{cap}, mem_j^{cap}\right)\) denoting the number of CPU cores, I/O bandwidth, network bandwidth and memory capacity of machine \(m_j\) respectively.

We define the set of applications as \(\boldsymbol{\Gamma} = \{ p_1, p_2, \cdots, p_i, \cdots , p_N \}\), where N is the number of applications, and $p_i$ represents the $i^{th}$ application. There can be numerous instances of a single application. The set of numbers of application instances is \(\boldsymbol{\eta} = \{ n_1, n_2, \cdots, n_i, \cdots , n_N \}\) and resource requirement for applications is \(\emph{\textbf{D}} = \{ d_1, d_2, \cdots, d_i, \cdots , d_N \}\), where $n_i$ and $d_i$ represents the number of instances and resource requirements of application $p_i$. Moreover, \(d_i\) is a four tuple \(\left(cpu_i^{req}, io_i^{req}, nw_i^{req}, mem_i^{req}\right)\) denoting the number of CPU cores, I/O bandwidth, network bandwidth and memory capacity required by application \(p_i\) respectively. All the instances of an application require the same amount of resources.

The users make application requests, and the set of application requests arrive at the cluster as shown in \autoref{fig:system-model}. The container mapper considers each container's resource requirements and machine affinity relationships while determining where to place its container instances. We consider the containers, once instantiated, to execute forever on the machine, similar to a web server or file server. In a machine, we may instantiate many instances of the same or many instances of different applications. The user request is a tuple containing information about user affinity preference (discussed later), the application service it requires and the number of instances of the application. We consider that each user has a fixed affinity preference. Each application instance initiates a container, and in our work, we efficiently place the container on the machines of the DC.

The container allocation matrix. i.e., the number of instances placed on each machine by different applications is represented as \(\emph{\textbf{B}} = \left[b_{ij}\right]_{N \times M}\), where \(b_{ij}\) is the number of instances of that has been placed on machine \(m_j\) of application \(p_i\).

We summarize all the symbols used in \autoref{tab:sys-mod-not} for ease of reading.

\begin{table}[]
\footnotesize
    \centering
    \begin{tabular}{|p{1.8cm}|p{6.2cm}|}
        \hline
        Symbol & Meaning \\
        \hline
        $M$, $N$ & Number of machines, applications \\
        $\boldsymbol{\Delta}$, $\boldsymbol{\Gamma}$ & Set of machines, applications \\
        $\emph{\textbf{V}}$ & Set of resource capacities of machine \\
        $v_j$ & Tuple of resources' capacity of $m_j$ \\
        $cpu_j^{cap}$ & CPU capacity of $m_j$ \\
        $\boldsymbol{\eta}$, $\emph{\textbf{D}}$ & Set of instances, resource required by applications \\
        $n_i$ & Number of instances of $p_i$ \\
        $d_i$ & Tuple of resources' required by $p_i$ \\
        $cpu_i^{req}$ & CPU required by $p_i$ \\
        $\emph{\textbf{B}}  \left[b_{ij}\right]_{N \times M}$ & Container Placement Matrix \\
        $b_{ij}$ & Number of instances of $p_i$ placed on $m_j$ \\
        $\emph{\textbf{S}} \left[s_{ij}\right]_{N \times M}$ & System affinity matrix \\
        $\beta_k$ & Affinity weights for k = 1, 2, 3 and 4 \\
        $\emph{\textbf{U}}\left[u_{ij}\right]_{N \times M}$ & User affinity matrix \\
        $\emph{\textbf{F}}  \left[f_{ij}\right]_{N \times M}$ & Affinity matrix \\
        $\bar{\emph{\textbf{F}}}  \left[\bar{f}_{ij}\right]_{N \times M}$ & Anti-Affinity matrix \\
        $\pi_j$, $P_j$ & Utilization of and power consumed by $m_j$ \\
        $P_j^{idle}$ & Idle power consumption and of $m_j$ \\
        $P_j^{max}$ & Maximum power consumption of $m_j$ \\
        $C^{power}$ & Total power cost \\
        $\phi_i^{affinity}$ & Total affinity payoff \\
        $C^{affinity}$ & Total affinity cost \\
        $\alpha$ & Affinity cost coefficient \\
        $C\left(\emph{\textbf{B}}\right)$ & Total system cost \\
        $\beta_1,\beta_2,\beta_3,\beta_4$& Affinity weights corresponding to CPU, I/O, network and memory resources\\
        $I$&Total number of instances of all the applications\\
        \hline 
    \end{tabular}
    \caption{Notations used}
    \label{tab:sys-mod-not}
\end{table}

\subsection{Affinity and Anti-affinity matrices}
\label{affinityantiaffinitysubsection}

The affinity matrix for a set of applications and machine nodes is given by \(\emph{\textbf{F}} = \left[f_{ij}\right]_{N \times M}\), where \(f_{ij}\) denotes the affinity of application \(p_i\) with machine node \(m_j\).

The values in the affinity matrix are obtained from the user affinity matrix and the system affinity matrix.

System affinity matrix  \(\emph{\textbf{S}} = \left[s_{ij}\right]_{N \times M}\) is calculated as follows:
\begin{equation}
    s_{ij} = 
    \left\{
        \begin{array}{l}
            0 \mbox{ , if } \left(cpu_j^{cap} < cpu_i^{req}\right) \mbox{ or } \left(io_j^{cap} < io_i^{req}\right) \mbox{ or } \\
            \quad \left(nw_j^{cap} < nw_i^{req}\right) \mbox{ or } \left(mem_j^{cap} < mem_i^{req}\right) \\
            \\
            \beta_1 \left(\frac{cpu_j^{cap} - cpu_i^{req}}{cpu_j^{cap}}\right) + \beta_2 \left(\frac{io_j^{cap} - io_i^{req}}{io_j^{cap}}\right) \\
            \quad + \beta_3 \left(\frac{nw_j^{cap} - nw_i^{req}}{nw_j^{cap}}\right) \\
            \quad + \beta_4 \left(\frac{mem_j^{cap} - mem_i^{req}}{mem_j^{cap}}\right) \mbox{ , otherwise}
        \end{array}
    \right.
    \label{eq:system-affinity-matrix}
\end{equation}

where \(\beta_1, \beta_2, \beta_3 \mbox{ and } \beta_4\) are affinity weights corresponding to CPU, I/O, network and memory resources respectively and \(\beta_1 + \beta_2 + \beta_3 + \beta_4 = 1\). Moreover, $cpu_j^{cap}$, $io_j^{cap}$, $nw_j^{cap}$, $mem_j^{cap}$ denotes the total number of CPU cores, maximum I/O bandwidth, maximum network bandwidth and maximum memory capacity of machine $m_j$. The system affinity matrix calculates affinity values because an application should have an affinity to a machine with better resources. It takes a real value in the range [0, 1].

As mentioned earlier, we also consider affinity values provided by users, who may prefer specific machines based on past usage experience, administrative reasons, or security concerns. Users can request services for multiple applications and multiple instances.

These values are in the form of a binary matrix called user affinity matrix \(\emph{\textbf{U}} = \left[u_{ij}\right]_{N \times M}\), where:
\begin{equation}
    {u}_{ij} = 
    \left\{
        \begin{array}{ll}
            1 & \mbox{, if machine \(m_j\) be affine to application \(p_i\)} \\
            0 & \mbox{, otherwise}
        \end{array}
    \right.
\end{equation}

The user specifies this matrix based on the affinity requirements. Now we calculate the final affinity matrix \(\emph{\textbf{F}} = \left[f_{ij}\right]_{N \times M}\) as follows:
\begin{equation}
    {f}_{ij} = \frac{{u}_{ij}+{s}_{ij}}{2}, \quad 0 \le f_{ij} \le 1
\end{equation}

Since $s_{ij}$ is a real number in the range [0, 1] and $u_{ij}$ is binary, calculated $f_{ij}$ values lie in the range of [0, 1].

The affinity parameters for a particular user are constant, so we treat each application-user pair as a separate application. For instance, if User X and User Y have different affinity preferences and both request service for the same application, we consider the application requests by User X and User Y as two distinct applications.

The anti-affinity factor rises when a particular application is not compatible with a certain machine or a certain machine does not have enough resources to run a certain application. The anti-affinity matrix is a binary matrix, given by \(\bar{\emph{\textbf{F}}} = \left[\bar{f}_{ij}\right]_{N \times M}\), where \(\bar{f}_{ij}\) is the anti-affinity relation of application \(p_i\) with machine node \(m_j\) and is determined as follows:
\begin{equation}
    \bar{f}_{ij} = 
    \left\{
        \begin{array}{ll}
            1 & \mbox{, if machine \(m_j\) does not affine to} \\
              & \mbox{   application \(p_i\)} \\
            0 & \mbox{, otherwise}
        \end{array}
    \right.
\end{equation}

It is to be noted that the value of $u_{i,j}$ and $\bar{f_{i,j}}$ can not be 1 for any values of $i$ and $j$. It is because since both are user-defined parameters and can not contradict each other, i.e. if $u_{i,j}=1$ then $\bar{f_{i,j}} \neq 1$ and if $\bar{f_{i,j}} = 1$ then $u_{i,j} \neq 1$.

It should be noted that affinity is a soft constraint, meaning that applications perform better when placed on machines they have an affinity with compared to machines they do not have an affinity with. Conversely, anti-affinity is a hard constraint, meaning applications cannot execute on machines with which they have anti-affinity. Anti-affinity specifies that an application should not be installed on a particular machine, making it a strict requirement.

\subsection{Power Consumption Model}
Reducing the overall power consumption usage in data centres is the primary objective of container placement problems \cite{yi2021, CHEUNG18}. As CPU consumption is the dominant factor influencing the power consumption of a machine, we use it to determine the power consumption. Other factors, such as I/O operations and memory usage, have a negligible impact on power consumption and are therefore ignored. The utilisation of the machine is the ratio of the number of CPU cores in the machine where the applications are running to the total number of CPU cores in the machine. The utilization \(\pi_j\) of machine \(m_j\) can, therefore, be expressed as a proportion of the number of CPU resources used currently to the total available CPU resources.

\begin{equation}
    \pi_j = \frac{\sum\limits_{i=1}^{N} cpu_i^{req} b_{ij}}{cpu_j^{cap}}
    \label{eq:utilization-exp}
\end{equation}

where $b_{ij}$ is the number of instances placed on machine $j$ by $i^{th}$ application. Considering the correlation between power used by the machine and CPU utilization, we consider the cubic power model as stated in \cite{cubic_power_model} to calculate the power consumed by a machine, which is given by:
\begin{equation}
    P_j = P_j^{idle} + \left(P_j^{max} - P_j^{idle}\right) \cdot \pi_j^3
    \label{eq:power-model}
\end{equation}

where \(P_j^{idle}\) and \(P_j^{max}\) are idle power and maximum power of machine \(m_j\) respectively. Therefore, the power consumption for the whole system is denoted by $C^{power}$ and can be expressed as:
\begin{equation}
    C^{power} = \sum\limits_{j=1}^{M}P_j
\end{equation}

\subsection{Affinity Payoff Model}
We defined the matrix \(\emph{\textbf{F}} = \left[f_{ij}\right]_{N \times M}\) to give the affinity relation of applications to the machine nodes in \autoref{affinityantiaffinitysubsection}. For application \(p_i\) on machine \(m_j\), we denote the expected affinity payoff as \(f_{ij}b_{ij}\), where $b_{ij}$ represents the number of instances placed by application $p_i$ on machine $m_j$. The total expected affinity payoff of application \(p_i\) on all the machines can, therefore, be defined as a convex combination as follows:
\begin{equation}
    \phi_i^{affinity} = \sum\limits_{j=1}^{M}f_{ij}b_{ij}
\end{equation}

A higher payoff means less cost. Thus, the total affinity cost can be written as negative of the total payoff of all applications since we want more applications to be placed on affinity machines, which can be expressed as:
\begin{equation}
    C^{affinity} = -\sum\limits_{i=1}^{N}\phi_i^{affinity}
\end{equation}

\subsection{Problem Formulation}
We propose our problem as an optimization problem. The objective is to minimize the total power along with maximizing the affinity payoff, and thus, the objective function can be written as:
\begin{equation*}
    \mbox{\emph{minimize }} C\left(\emph{\textbf{B}}\right) = C^{power} + \alpha . C^{affinity}
\end{equation*}
\begin{equation}
    \begin{aligned}
        C\left(\emph{\textbf{B}}\right) &= \sum\limits_{j=1}^{M}\left(P_{j}^{idle} + \left(P_j^{max} - P_j^{idle}\right) {\left(\frac{\sum\limits_{i=1}^{N} cpu_i^{req} b_{ij}}{cpu_j^{cap}}\right)}^{3} \right) \\
        &\quad - \alpha\sum\limits_{j=1}^{M} \sum\limits_{i=1}^{N}f_{ij}b_{ij} \\
        &=\sum\limits_{j=1}^{M}\left(\left(P_j^{max} - P_j^{idle}\right) {\left(\frac{\sum\limits_{i=1}^{N} cpu_i^{req} b_{ij}}{cpu_j^{cap}}\right)}^{3} \right) \\
        &\quad - \alpha\sum\limits_{j=1}^{M} \sum\limits_{i=1}^{N}f_{ij}b_{ij} + \sum\limits_{j=1}^{M}P_j^{idle} \\
    \end{aligned}
    \label{eq:power-exp}
\end{equation}

The primary goal is to minimize the system's overall power consumption with minimization of the affinity cost incorporated into the objective function based on this. Consequently, we define a variable cost coefficient \(\alpha\) to highlight the significance of affinity cost optimisation. We can adjust the value of \(\alpha\) to achieve the required trade-off between power and affinity.

Since \(\sum\limits_{j=1}^{M} P_j^{idle}\) is a constant term and does not affect the solution of the original expression, we can remove it and re-write new optimization function as:
\begin{equation}
    \begin{aligned}
        \bar{C}\left(\emph{\textbf{B}}\right) &= \sum\limits_{j=1}^{M}\left(\left(P_j^{max} - P_j^{idle}\right) {\left(\frac{\sum\limits_{i=1}^{N} cpu_i^{req} b_{ij}}{cpu_j^{cap}}\right)}^{3} \right) \\
        &\quad - \alpha\sum\limits_{j=1}^{M} \sum\limits_{i=1}^{N}f_{ij}b_{ij}
    \end{aligned}
    \label{eq:power-exp-new}
\end{equation}
Following is the list of constraints that are addressed while optimizing \autoref{eq:power-exp-new}.
\begin{itemize}
    
\item{\textbf{Anti-Affinity Constraint:}}
We defined the anti-affinity matrix \(\bar{\emph{\textbf{F}}} = \left[\bar{f}_{ij}\right]_{N \times M}\) to check if an application has an anti-affinity relationship to a machine node. The application \(p_i\) is not deployed on machine \(m_j\), if \(\bar{f}_{ij} = 1\). Hence, it is taken as a strict constraint and can be expressed as:
\begin{equation}
    \sum\limits_{j=1}^{M} \sum\limits_{i=1}^{N} \bar{f}_{ij}b_{ij} = 0
    \label{eq:anti-affinity-constraint}
\end{equation}

\item{\textbf{Application Allocation Constraint:}}
This constraint signifies that the user request is completely satisfied, i.e., all the instances of all applications are deployed on some machine server. This can be expressed as follows:
\begin{equation}
    \sum\limits_{j=1}^{M}b_{ij} = n_{i}  \mbox{, } \forall 1 \leq i \leq N
    \label{eq:app-all-constraint}
\end{equation}

\item{\textbf{Resource Capacity Constraint:}}
Each server node has limited resource capacities which can not be exceeded. Thus, the allocation of application instances must be within these limits. This constraint can be written as:
\begin{equation}
    \begin{aligned}
        \sum\limits_{i=1}^{N}d_{i}b_{ij} \leq v_{j} \mbox{, } &\forall 1 \leq j \leq M \mbox{, } \\
        & d_i = \left(cpu_i^{req}, io_i^{req}, nw_i^{req}, mem_i^{req}\right) \mbox{, } \\
        & v_j = \left(cpu_j^{cap}, io_j^{cap}, nw_j^{cap}, mem_j^{cap}\right)
    \end{aligned}
    \label{eq:resource-cap-constraint}
\end{equation}
\end{itemize}
Considering equations \autoref{eq:power-exp-new}, \autoref{eq:anti-affinity-constraint}, \autoref{eq:app-all-constraint} and \autoref{eq:resource-cap-constraint}, we can express the final container allocation problem as an integer non-linear optimization problem.

In this problem, we map applications to machines, considering multiple types of resources. This mapping can be formulated as a 0/1 Multi-Objective Multidimensional Knapsack Problem (MKP), which is known to be NP-complete \cite{Laabadi2018}. Additionally, we incorporate the cubic power model \cite{cubic_power_model} and an affinity factor to minimize the combined cost. We propose a heuristic placement strategy as the cubic power model cannot be solved using Integer Linear Programming (ILP) methods. The challenge with the cubic power model arises from the non-linear increase in power consumption with the computational load. Mathematically, it can be demonstrated that the total power consumption cost is minimized when all machines operate with equal load. However, this approach increases the affinity cost, presenting a significant challenge that our work addresses.

\section{Solution Approach}

We proposed three solution approaches to address the problem mentioned in the previous section. The approaches are: (a) Power Aware Container Placement (PAP): This placement strategy aims to minimize power consumption costs. (b) Affinity Aware Container Placement (AAP): This approach prioritizes minimizing the affinity cost. (c) Combined Power and Affinity Container Placement (CPAAP): This approach combines the strategies of PAP and AAP. Each of these approaches is explained in detail below.

\begin{figure}
    \centering
    \includegraphics[scale=0.58]{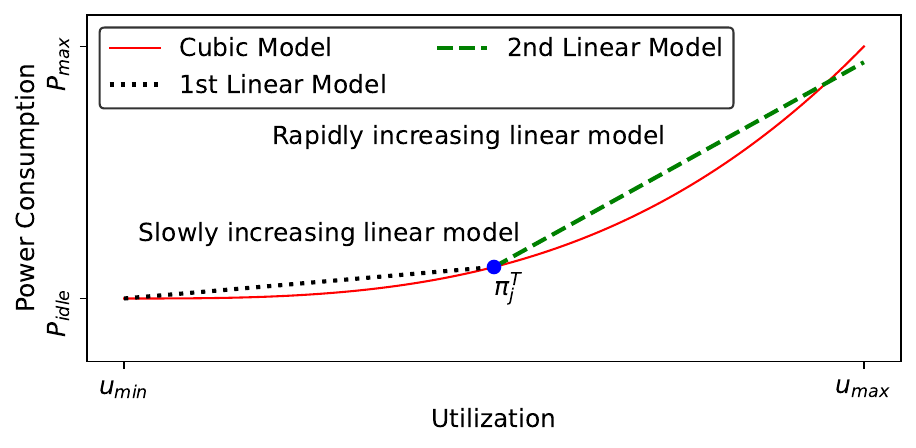}
    \caption{\footnotesize Approximation of cubic power model using two linear models. $\pi_j^T$ denotes the splitting points of two linear models.}
    \label{fig:cubic-linear-fit}
\end{figure}

A key consideration in our problem statement is that the idle power term $\sum\limits_{j=1}^{M} P_j^{idle}$ remains constant and does not affect the value of our optimization function. Therefore, saving some machines does not provide any power savings since they still consume power while idle. Instead, utilizing all available machines to serve user requests is more effective. This approach helps distribute the workload more evenly, reducing the utilization of individual machines, decreasing their power consumption, and ultimately lowering the total system cost.

\subsection{Power Aware Container Placement (PAP)}

We design our first approach as shown in \autoref{alg:greedy-power} considering only the power aspect to minimize the total system cost. \autoref{alg:ini_param} initializes the parameters. The final affinity matrix $\emph{\textbf{F}}$ is derived from system affinity matrix ($\emph{\textbf{F}}$) and user affinity matrix ($\emph{\textbf{B}}$). 
Our starting point is an empty allocation matrix $\emph{\textbf{B}}$. In the beginning, we arrange the applications according to descending resource requirements.

The power model described in \autoref{eq:power-model} is a cubic model. This cubic model can be approximated as a combination of two linear models, as depicted in \autoref{fig:cubic-linear-fit}. These linear models are obtained by fitting a linear approximation to different cubic curve segments. Power consumption increases slightly with respect to utilization in the first linear approximation, while the second linear approximation increases rapidly. We denote this split between the two linear models with $\pi_j^T$ for machine $j$. The optimum value of $\pi_j^T$ varies for different machines depending on the value of $P^{idle}_j$ and $P^{max}_j$. Most of the realistic values of $P^{idle}_j$ and $P^{max}_j$ suggest the value of $\pi^T_j$ in the range of 0.4 to 0.6 \cite{piT}.

We initiate a priority queue of machines, where the initial value of the priority parameter ($\omega$) for all machines is set to zero. The first machine in this priority queue with adequate remaining resources and does not have anti-affinity with the application is selected for placement. Let us consider that machine $m_j$ is selected. If the machine's utilization is less than $\pi_j^T$, the priority parameter is updated to the current utilization of the machine (\emph{i.e.} $\omega_j=\pi_j^T$). If the priority parameter is more than $\pi_j^T$ but less than 1, then the priority parameter is set to one (\emph{i.e.} $\omega_j=1$). If the selected machine's priority parameter is one or greater, it is doubled (\emph{i.e.} $\omega_j=2 \times \omega_j$). This adjustment ensures that machines with utilization exceeding $\pi_j^T$ are prioritised less. We set the machine's priority parameter to one and double it after every access to ensure the number of placements after the machine's utilization crosses $\pi_j^T$ is minimized. This process is iterative and continues until no applications are left for placement. The primary goal of PAP is to balance the CPU load as efficiently as possible.

\begin{algorithm}
    \caption{Initialize Parameters}
    \label{alg:ini_param}
    \footnotesize
    \KwIn{Affinity matrix $\emph{\textbf{B}}$, anti-affinity matrix $\bar{F}$, set of applications $\Gamma$}
    \KwOut {Affinity matrix $\emph{\textbf{F}}$, container allocation matrix $\emph{\textbf{B}}$}
    $S \gets$ Calculate system affinity matrix using \autoref{eq:system-affinity-matrix} \\
    $F \gets \frac{1}{2}  * (U + S)$ \\
    $B \gets 0$, Initialize allocation matrix with all zero \\
    Sort $\Gamma$ in decreasing order of resource requirement following the order $\left(cpu, io, nw, mem\right)$ \\
    
\end{algorithm}

\begin{algorithm}
\caption{Power Aware Placement (PAP)}\label{alg:greedy-power}
\footnotesize

\KwIn {Set of machines $\Delta$, applications $\Gamma$, User affinity matrix $\emph{\textbf{B}}$, Anti-affinity matrix $\bar{F}$, Lambda vector $\Lambda$}
\KwOut{Container allocation matrix $\emph{\textbf{B}}$}
Initialize Parameters using \autoref{alg:ini_param} \\
$\pi_j \gets 0 \mbox{ } \forall \mbox{ } 0 \le j \le M$ \\
$\omega_j \gets 0$, Initialize Priority Queue parameter for machines \\
$Q \gets \Delta$, Initialize a Priority Queue based on the parameter $\omega$. Lower value of $\omega$ is given higher priority \\
\For{$p_i$ in $\Gamma$}
    {\For{$k$ in range($n_i$)}
        {$m_j \gets $ Find first-best machine satisfying $\bar{f_{ij}} = 0$ and having available resources using $Q$ \\
        $b_{ij} \gets b_{ij} + 1$, Place $k^{th}$ instance of $p_i$ on $m_j$ \\
        Update utilization $\pi_j$ of $m_j$ \\
        
        \eIf{$\omega_j < \pi_j^T$}
        {
            $\omega_j \gets \pi_j$
        }
        {
            \eIf{$\omega_j < 1$}
            {
                $\omega_j \gets 1$
            }
            {
                $\omega_j \gets 2 \times \omega_j$
            }
        }
        }
    } 
    
\Return $\bm{B}$
\end{algorithm}

\subsection{Affinity Aware Container Placement (AAP)}
The second aspect of our optimization problem is the affinity between machines and applications. The more the applications are placed on more affine machines, the more it reduces the system cost. In order to find a solution based on this idea, we can design a greedy approach to place applications on machines having the highest affinity as mentioned in \autoref{alg:greedy-affinity}. Initially, we arrange the applications in decreasing order of needed resources. Then, the container mapper looks for the machine with the highest affinity for the current application instance and sufficient resources to place the instance. Hence, the container mapper selects the machine with the higher affinity in every instance out of greed. Following this, we build up our allocation matrix $\emph{\textbf{B}}$ as our final result. This approach does not consider load balancing.

\begin{algorithm}[tb!]
\caption{Affinity Aware Placement (AAP)}\label{alg:greedy-affinity}
\footnotesize

\KwIn{Set of machines $\Delta$, applications $\Gamma$, User affinity matrix $\emph{\textbf{B}}$, Anti-affinity matrix $\bar{F}$}
\KwOut{Container allocation matrix $\emph{\textbf{B}}$}
Initialize Parameters using \autoref{alg:ini_param} \\

\For{$p_i$ in $\Gamma$}
    {
    
    \For{$k$ in range($n_i$)}
        {
        $m_j \gets argmax_j \left(f_{ij}\right)$, find machine satisfying $\bar{f_{ij}} = 0$ and having available resources \\
        
        $b_{ij} \gets b_{ij} + 1$, Place $k^{th}$ instance of $p_i$ on $m_j$
        }
    }

\Return $\bm{B}$

\end{algorithm}

\subsection{Combine Power and Affinity Aware Container Placement (CPAAP)}
In summary, both approaches are good in some cases, but to get to a better solution for our problem in all cases, we need to consider both aspects of our problem simultaneously. The problem with PAP is that it does not consider the affinity payoff of applications. If some machines have a high affinity for an application but their utilization is also high, they are not selected for placement of applications. However, it can be the case that choosing those machines might be more profitable. The AAP brings affinity to the scenario but does not consider the utilization of machines, which may lead to some machines using more power, resulting in more system costs. Hence, we need a solution between the two extremes and consider both the utilization and affinity of machines together. 

We design such an approach in \autoref{alg:main}. Initially, we arrange the applications according to progressively lower resource requirements. We start a priority queue of machines, where the machines are arranged in increasing order of utilisation. Next, we continue iterating over each application instance, identifying two machines for each instance, $j_1$ and $j_2$. Out of all the machines having enough remaining resources to place the current instance and having no anti-affinity constraint ($\bar{f_{ij}} = 0$), say machine $j_1$ is the one having lowest current utilization and assume machine $j_2$ has the highest affinity for the application (\emph{i.e.} $argmax_j \left(f_{ij_2}\right)$). Then, we calculate the change in current system cost if the container mapper places the application on either machine ($cost_1$ and $cost_2$ respectively). To calculate the increase in system cost, we can simply use the following equation.

\begin{equation}
    c_{ij} = (P_j^{max} - P_j^{idle})((\pi_j^{new})^3 - (\pi_j^{old})^3) - \alpha f_{ij}
    \label{eq:cost-change}
\end{equation}

where, $c_{ij}$ is increase in total system cost for placement application $i$ on machine $j$. $\pi_j^{new}$ and $\pi_j^{old}$ are the new and old utilization of machine $j$, i.e., before and after placement of application $i$ on it respectively. The $P_j^{idle}$ term does not affect the increase in system cost. Hence, it can simply be calculated by taking the difference of utilization cube multiplied by $(P_j^{max} - P_j^{idle})$ and subtracting the affinity payoff $\alpha f_{ij}$ from it. If $cost_1$ is less than $cost_2$, then machine $j_1$ is selected, and the current instance is placed on it. Else, we go with machine $j_2$. At last, $\emph{\textbf{B}}$ is given as output, depicting a final solution that would minimize the total cost of the system.

\begin{algorithm}[tb!]
\caption{Combine Power and Affinity Aware Placement (CPAAP)} \label{alg:main}
\footnotesize

\KwIn{ Set of machines $\Delta$, applications $\Gamma$, Affinity matrix $\emph{\textbf{B}}$, Anti-affinity matrix $\bar{F}$}

\KwOut{Container allocation matrix $\emph{\textbf{B}}$}

Initialize Parameters using \autoref{alg:ini_param} \\
$\pi_j \gets 0 \mbox{ } \forall \mbox{ } 0 \le j \le M$ \\
$Q \gets \Delta$, Initialize a Priority Queue based on the utilization $\pi_j$. Lower value of $\pi$ is given higher priority \\
\For{$p_i$ in $\Gamma$}
    {\For{$k$ in range($n_i$)}
        {$m_{j_1} \gets$ Find machine satisfying 
        $\bar{f_{ij_1}} = 0$ and having available resources using $Q$\\
        $cost_1 \gets$ Calculate increase in cost if 
        $p_i$ is placed on $m_{j_1}$ using \autoref{eq:cost-change} \\
        $m_{j_2} \gets argmax_{j_2} \left( f_{ij_{2}} \right)$, find machine satisfying $\bar{f_{ij_{2}}} = 0$ and having available resources \\
        
        $cost_2 \gets$ calculate increase in cost if $p_i$ is placed on $m_{j_{2}}$ using \autoref{eq:cost-change} \\
        
        \eIf{$cost_1 \le cost_2$}
            {$j \gets j_1$}
            {$j \gets j_2$}
        
        $b_{ij} \gets b_{ij} + 1$, Place $k^{th}$ instance of $p_i$ on $m_j$ \\
        Update utilization $\pi_j$ of $m_j$ \\
        }
    }
\Return $\bm{B}$

\end{algorithm}

\subsection{Computational Complexity}
We define the term $I$ to denote the total number of instances of all applications. Therefore, we can express this as $I = \sum\limits_{i=1}^N n_i$.

For the PAP and AAP approaches, the container mapper identifies a suitable machine for each instance of the applications sequentially, resulting in a complexity of $O(IM)$.

In the case of CPAAP, the container mapper identifies two machines for each instance of the applications sequentially: one machine that consumes the least power and another that has the highest affinity with the application. From these two machines, the container mapper selects the better one, which gives lower system cost. Consequently, the complexity for CPAAP is $O(I(M + M))$, which simplifies to $O(IM)$.

\section{Experimental Setup and Results}
In this section, we present simulation results to evaluate the performance of our proposed approaches.

\subsection{Simulation Setup}
We developed a homegrown simulator, similar to \cite{CloudSim} and simulated experiments. We implemented the simulation environment using C++ and Python and conducted experiments comparing our approach with other state-of-the-art approaches on synthetic and real-life datasets. Furthermore, we define preference relationships for the machines, i.e., affinity and anti-affinity matrices. 

\subsubsection{Synthetic Dataset} 
We determine the values of the resource capacities of machines, the value of idle and maximum power, the resource requirements of applications, and the number of instances of each application using random distribution and a certain predefined range for each parameter. We generate system affinity values from the generated values using \autoref{eq:system-affinity-matrix} and user affinity matrix at random. We generated an anti-affinity matrix by defining a fixed percentage of machines; each application has anti-affinity. 

\subsubsection{Real-life Dataset} We also experimented on a real-life dataset, namely Google Cluster \cite{clusterdata:Wilkes2011}. Additionally, the parameters missing from the dataset are again generated from normal distribution defining a mean and standard deviation. We extracted traces from these datasets and incorporated affinity and anti-affinity matrices as described in the previous methods.

To perform our simulation, we generated traces of 200 applications and 250 machines, each with 1-4 instances, resulting in approximately 600 application instances.

\subsection{Metrics}
We evaluate the performance of our proposed approach using four aspects: total system cost, average affinity satisfaction ratio, average utilization and affinity payoff ratio of machines. 
\subsubsection{Total System Cost} The total cost incurred by the system upon placing the application request on the machine cluster is calculated using \autoref{eq:power-exp}. The lower the total system cost, the better the allocation of containers.
\subsubsection{Average Affinity Satisfaction Ratio} The average proportion of application containers placed on the machines it is affine to, denoted by $\rho$. It can be calculated by taking element-wise multiplication of two matrices: User affinity matrix $\emph{\textbf{U}}$ and container allocation matrix $\emph{\textbf{B}}$ and then summing up all the values, divided by total number of application instances. The higher the ratio value, the better the allocation of requests on cluster machines.
\begin{equation}
    \rho = \frac{\sum\limits_{j=1}^{M} \sum\limits_{i=1}^{N} u_{ij} b_{ij}}{\sum\limits_{i=1}^{N} n_i}
\end{equation}
\subsubsection{Average Utilization} The average utilization of all machines after completely placing application requests, where utilization of a machine is calculated using \autoref{eq:utilization-exp}.
\subsubsection{Affinity Payoff Ratio} The ratio of affinity payoff to the total system cost, denoted by $\psi$. The higher value of the ratio indicates better affinity satisfaction and minimizes the power consumption of the cluster.
\begin{equation}
    \psi = \frac{\sum\limits_{i=1}^{N}\phi_i^{affinity}}{C\left(\emph{\textbf{B}}\right)}
\end{equation}
\begin{figure*}[tb!]
    \includegraphics[scale=0.25]{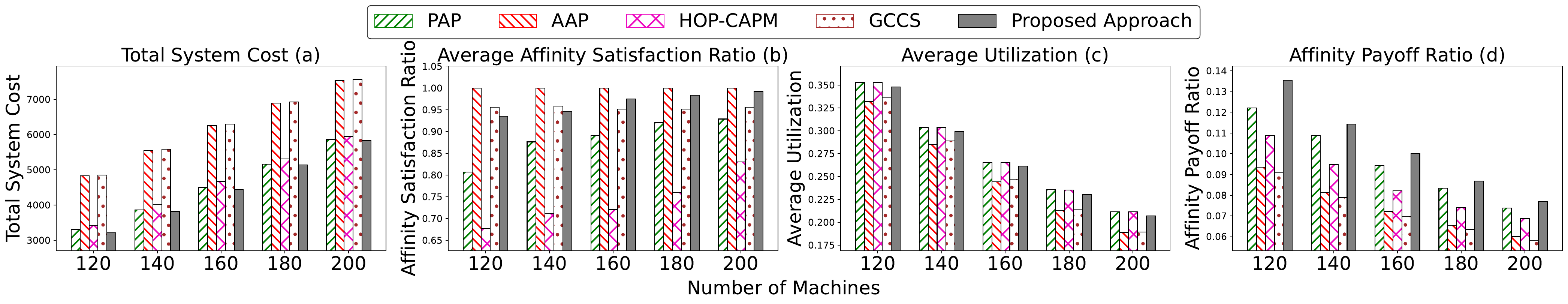}
    \caption{\footnotesize Results of experiment with 175 applications and different number of machines on real-life dataset}
    \label{fig:mach-bench}
\end{figure*}
\begin{figure*}
    \includegraphics[scale=0.25]{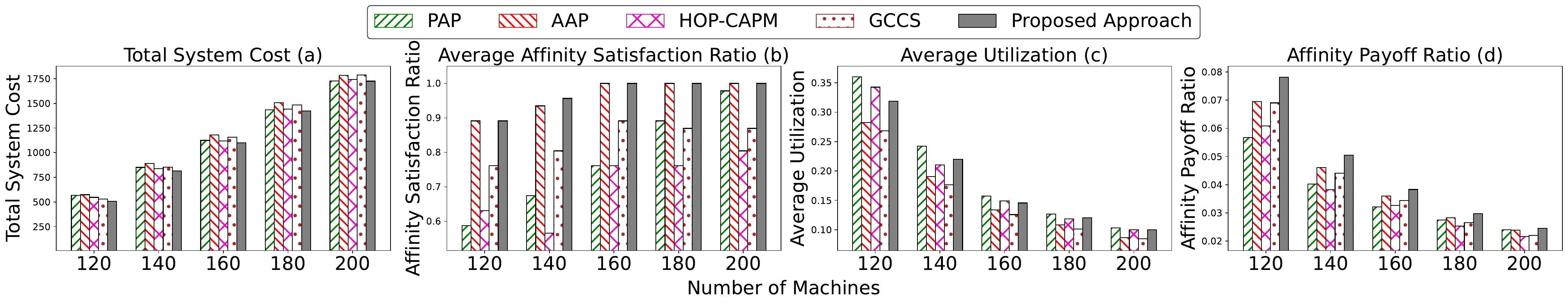}
    \caption{\footnotesize Results of experiment with 20 applications and different number of machines on synthetic dataset}
    \label{fig:mach-random}
\end{figure*}
\subsection{State-of-the Art Algorithms}
We compare the proposed approach with two state-of-the-art heuristic placement approaches: Global Cost-Aware Container Scheduling Algorithm (GCCS) \cite{global_cost_aware} and Heuristic of Placement for Constraint Aware Profit Maximization problem (HOP-CAPM) \cite{group-scheduling}. The specific explanation of the two approaches is given below.
\subsubsection{GCCS} It models the number of containers per server selected by the application as an integer linear program (ILP) by using a linear power model. It attempts to find a sub-optimal solution of the ILP by fixing the solution to a relaxed LP version of the same problem \cite{global_cost_aware}. We examine the performance of this solution of the linear power model on cubic power model.
\subsubsection{HOP-CAPM} In order to minimize the overall system cost of the cluster, this approach takes into account the estimation of task execution time in a heterogeneous environment, effective task ordering and profit-based job allocation. The approach first divides the machines and applications into several groups based on the hard constraints and then places each group of applications separately by selecting a machine group based on certain conditions \cite{group-scheduling}. As our problem considers web server applications which do not have any deadlines, we modified the HOP-CAPM approach without considering deadlines and placed the applications on machines by dividing them into groups based on hard constraints.

\subsection{Simulation Results}
In this section, we conduct several experiments and compare the performance of the proposed approach (Algorithm \ref{alg:main}) with state-of-the-art approaches, GCCS and HOP-CAPM along with PAP (Algorithm \ref{alg:greedy-power}) and AAP (Algorithm \ref{alg:greedy-affinity}) as well on the metrics mentioned in the previous section. The value of $\beta_1$, $\beta_2$, $\beta_3$ and $\beta_4$, used to calculate the system affinity matrix in \autoref{eq:system-affinity-matrix}, is fixed to be 0.4, 0.2, 0.2 and 0.2 respectively for all the experiments. Also, the value of affinity cost coefficient $\alpha$ is fixed to be 4 (as considered in \cite{global_cost_aware}) for the experiments unless specified otherwise.

\begin{figure*}[tb!]
    \includegraphics[scale=0.25]{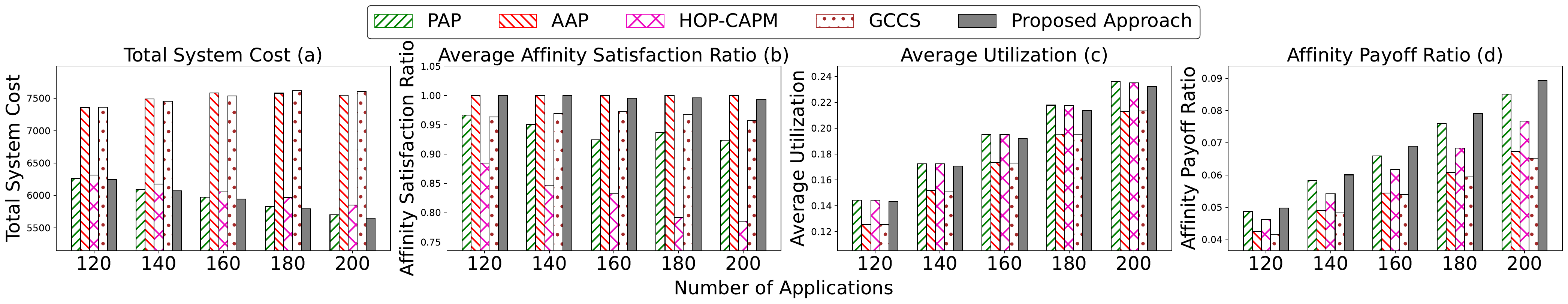}
    \caption{\footnotesize Results of experiment with 250 machines and different number of applications on real-life dataset}
    \label{fig:app-bench}
\end{figure*}

\begin{figure*}
    \includegraphics[scale=0.25]{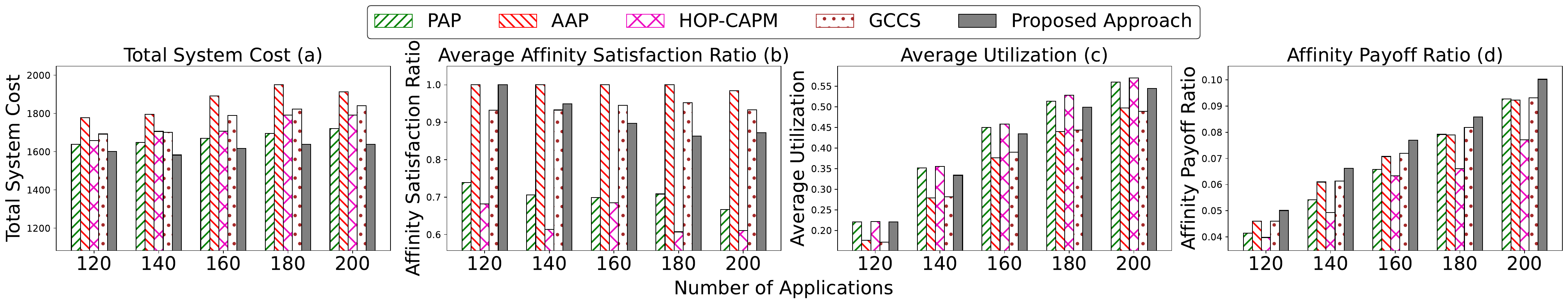}
    \caption{\footnotesize Results of experiment with 60 machines and different number of applications on synthetic dataset}
    \label{fig:app-rand}
\end{figure*}

\begin{figure*}
    \centering
    \includegraphics[scale=0.25]{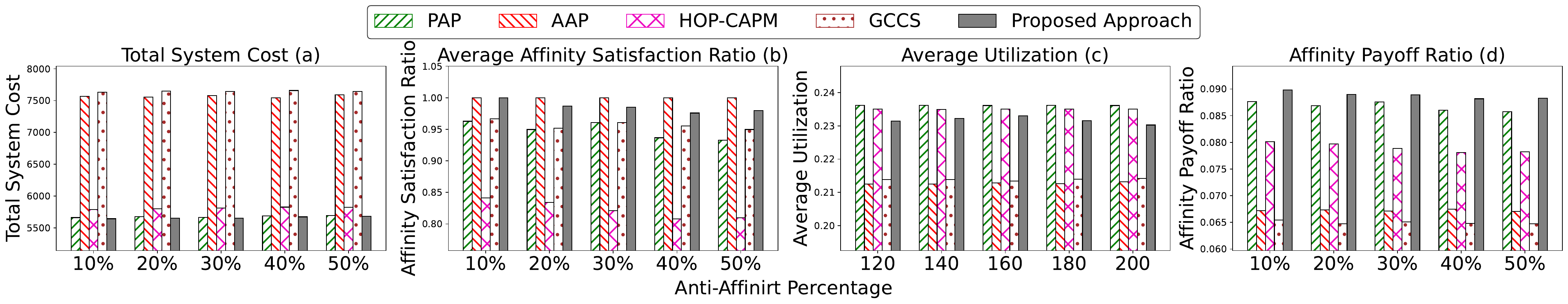}
    \caption{\footnotesize Results of experiment with varying anti-affinity percentage for 250 machines and 200 applications on real-life dataset}
    \label{fig:antiaffinity}
\end{figure*}

\subsubsection{The Impact of Number of Machines}
In this experiment, we fix the number of applications while varying the number of machines available. We fix the number of applications to 175 and 20 for real-life and synthetic datasets, respectively.

\autoref{fig:mach-bench} (a) shows the total system cost by each approach with 175 applications on different numbers of machines on real-life datasets. We can clearly see that our proposed approach performs better than all the other approaches, giving slightly better results than HOP-CAPM while completely outperforming GCCS. This is because our proposed approach focuses on optimizing both power cost and affinity payoff, leading to better results. Further, GCCS provides a solution which tries to optimize a linear power cost model. When we use the same solution with the cubic power model, it does not perform well since the single linear model is not a very good approximation of a cubic power model, and thus, there might be better solutions than the one given by GCCS. PAP achieves a lower total cost by maintaining a low power consumption, but still our approach performs better since we consider both affinity satisfaction and power consumption. AAP does not perform that for total cost minimization since it does not consider power consumption and only aims to achieve affinity satisfaction.

We can also see that our approach gives better results in terms of affinity satisfaction of applications as shown in \autoref{fig:mach-bench} (b), beating PAP and HOP-CAPM by large margins while reaching as good as GCCS, beating even it for higher number of machines. However, it cannot defeat AAP (though fairly close enough) since AAP focuses only on affinity satisfaction, thus achieving the highest affinity satisfaction each time. Further, as depicted in \autoref{fig:mach-bench} (c), the proposed approach utilizes all machines evenly reaching as good average utilization levels as PAP, which only focuses on the utilization of machines for placement. \autoref{fig:mach-bench} (d) shows that our approach always performs better than other approaches in terms of affinity payoff ratio. It indicates that our approach gives better affinity satisfaction under similar power consumption. Although PAP and the proposed approach look equally appealing in terms of total system cost, this graph shows that our approach achieves better affinity satisfaction by not spending much more power than PAP and, at the same time, also maintaining high affinity payoffs. Similar results can also be seen for the synthetic dataset as shown in \autoref{fig:mach-random}, where 20 applications are placed on different machines.

\subsubsection{The Impact of Number of Application Requests}
In this experiment, we fix the number of machines while varying the number of applications placed on them. Similar to the previous experiment, we fix the number of machines to 250 and 60 for real-life and synthetic datasets, respectively.

\autoref{fig:app-bench} (a) shows the total system cost by each approach with 250 machines for different numbers of application requests on real-life datasets. With the increase in the number of applications, the power cost of all approaches shows an increasing trend. This is because the load in the cluster increases with an increase in application requests, and the container mapper places them on the same number of machines, increasing power usage. We see that our proposed approach achieves lower total system cost as the number of machines increases. It is because the load is balanced nicely with an abundance of machines. Again, we can clearly see that our proposed approach beats all the other approaches, giving slightly better results than HOP-CAPM and PAP while completely outperforming GCCS and AAP. This is because our proposed approach considers a global perspective while balancing power and affinity, leading to better results.

Further, we can see it gives better results in terms of affinity satisfaction of applications as shown in \autoref{fig:app-bench} (b), beating HOP-CAPM by a large margin while reaching as good as AAP. However, it cannot defeat AAP since AAP focuses only on affinity satisfaction, thus achieving complete affinity satisfaction most of the time. This time, our approach beats even GCCS for all the cases. Further, as depicted in \autoref{fig:app-bench} (c), the proposed approach utilizes all machines evenly reaching as good average utilization levels as PAP, which focuses only on the utilization aspect for placement.\autoref{fig:app-bench} (d) shows that our approach always performs best in terms of affinity payoff ratio. Also, we see that the payoff ratio decreases as the load decreases, which explains the decrease in total system cost with increasing load due to the fact we are getting better affinity payoffs. Again, we see similar results for the synthetic dataset as shown in \autoref{fig:app-rand}, where we used 60 machines with different numbers of applications.

\subsubsection{The Impact of Anti-Affinity Percentage}
Anti-affinity percentage is the average number of machines the user specifies per application to be anti-affine using anti-affinity matrix $\bar{\emph{\textbf{F}}}$. We consider real-life datasets and take different anti-affinity matrices such that the anti-affinity percentage changes from 10\% to 50\%, increasing 10\% at a time. This analysis helps in analysing performance under an increasingly constrained environment. More anti-affinity percentage leads to stricter constraints and gives the container mapper less freedom to place instances. This makes it difficult to fulfil affinity satisfaction.

As depicted by \autoref{fig:antiaffinity}, even increasing the percentage of anti-affinity machines does not hamper the performance of our approach. Increasing the anti-affinity percentage from 10\% to 50\% shows a minimal increase in total system cost and a minimal decrease in affinity satisfaction as compared to other approaches as shown in \autoref{fig:antiaffinity} (a) and \autoref{fig:antiaffinity} (b) respectively. Further, \autoref{fig:antiaffinity} (c) depicts that the average utilization also does not get much affected by an increase in anti-affinity percentage. \autoref{fig:antiaffinity} (d) again shows that our proposed approach has the best affinity payoff ratio depicting in all situations, it keeps good affinity satisfaction and lower power consumption, both simultaneously, outperforming all the other approaches.

\begin{figure}
    \centering
    \includegraphics[scale=0.53]{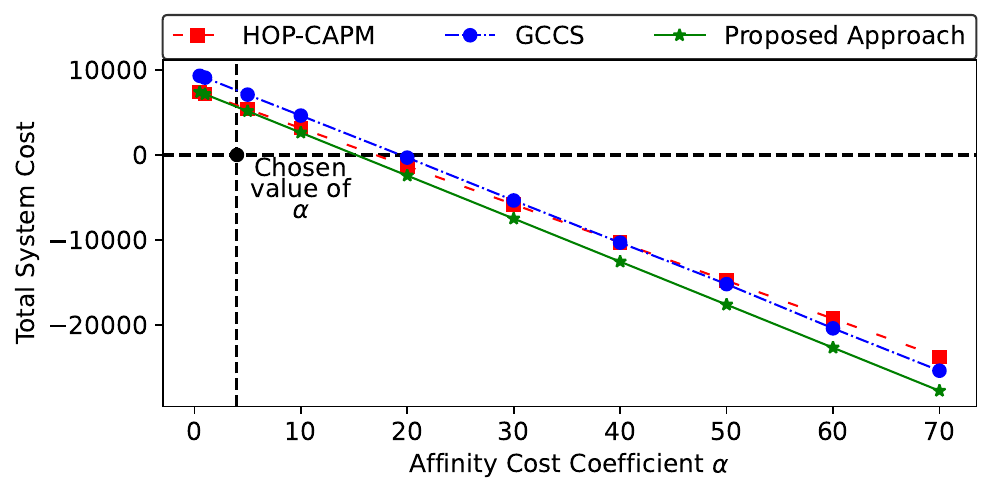}
    \caption{\footnotesize Impact of affinity cost coefficient $\alpha$ on total system cost for real-life dataset with 200 applications and 250 machines}
    \label{fig:imp-alpha}
\end{figure}
\subsubsection{The Impact of Affinity Cost Coefficient \texorpdfstring{$\alpha$}{Lg}}
For this experiment, we only consider the real-life dataset. We take all 200 applications to be placed on 250 machines while varying the value of $\alpha$ in the range of 0.5 to 70 and calculating the total system cost for each value.

\autoref{fig:imp-alpha} shows the change in total system cost for a range of affinity cost coefficient ($\alpha$) values. For small values of $\alpha$, HOP-CAPM and our proposed approach give almost similar results (though our proposed approach beats them by small margins). This is because for smaller values of alpha, only the power cost minimization matters, and affinity satisfaction is not that important, so minimizing power through PAP gives good results, too. But as we increase the value of $\alpha$, all approaches eventually beat HOP-CAPM. Meanwhile, our proposed approach always gives better results no matter what $\alpha$ value we took. The variation in total system cost for PAP and AAP is not shown in \autoref{fig:imp-alpha} due to space constraints. However, both approaches exhibit similar trends, and our proposed approach outperforms them both.

\section{Conclusion}
We designed a Combine Power and Affinity Aware Placement approach to solve the cost minimization problem of container allocation in heterogeneous server clusters. We considered multiple resources of machines and containers and calculated the total cost of the system, aiming at saving the total power consumption of the system and trying to meet the affinity of the application. We proposed a heuristic in which the container mapper places a container on one of the two machines(one with the lowest current utilization and one with the most affinity for the current container), which leads to the lowest current total cost increase. We studied the impact on the performance of the proposed heuristic due to changing system load by varying the number of machines and applications. We also examined the effect of making the system more constrained by varying the anti-affinity percentage. Then, we analysed the effect of varying the affinity cost coefficient.

%The datasets used, along with the implementation details of our approach, are available at \cite{github_link} for your reference.

%\newpage
\footnotesize
\setstretch{0.7}
%\bibliography{btp}

\begin{thebibliography}{10}

\bibitem{yi2021}
C.~Yi, R.~Dou, D.~Chen, M.~Guo, and J.~Wang.
\newblock National integrated big data center collaborative innovation system: Overall framework and strategic value.
\newblock {\em E-Government}, (06):2--10, 2021.

\bibitem{CHEUNG18}
Howard Cheung, Shengwei Wang, Chaoqun Zhuang, and Jiefan Gu.
\newblock A simplified power consumption model of information technology (it) equipment in data centers for energy system real-time dynamic simulation.
\newblock {\em Applied Energy}, 222:329--342, 2018.

\bibitem{global_cost_aware}
Saiqin Long, Wen Wen, Zhetao Li, Kenli Li, Rong Yu, and Jiang Zhu.
\newblock A global cost-aware container scheduling strategy in cloud data centers.
\newblock {\em IEEE Transactions on Parallel and Distributed Systems}, 33(11):2752--2766, 2022.

\bibitem{kubernetes}
Kubernetes.
\newblock {\em Kubernetes: Production-grade container orchestration}, June 2021.
\newblock Accessed: Nov, 2023.

\bibitem{docker}
Dirk Merkel.
\newblock Docker: Lightweight linux containers for consistent development and deployment, 2014.

\bibitem{moreno2018}
R.~Moreno-Vozmediano, R.~S. Montero, E.~Huedo, and I.~M. Llorente.
\newblock Orchestrating the deployment of high availability services on multi-zone and multi-cloud scenarios.
\newblock {\em Journal of Grid Computing}, 16(1):39--53, 2018.

\bibitem{monitoring_audit_logging}
Philippe Massonet, Syed Naqvi, Christophe Ponsard, Joseph Latanicki, Benny Rochwerger, and Massimo Villari.
\newblock A monitoring and audit logging architecture for data location compliance in federated cloud infrastructures.
\newblock In {\em IEEE Int. Symp. on Parallel and Distributed Processing}, pages 1510--1517, 2011.

\bibitem{espling2016}
D.~Espling, L.~Larsson, W.~Li, J.~Tordsson, and E.~Elmroth.
\newblock Modeling and placement of cloud services with internal structure.
\newblock {\em IEEE Transactions on Cloud Computing}, 4(4):429--439, Oct.--Dec. 2016.

\bibitem{group-scheduling}
Chinmaya Swain, Bhawana Gupta, and Aryabartta Sahu.
\newblock Constraint aware profit maximization scheduling of tasks in heterogeneous datacenters.
\newblock {\em Computing}, 102, 10 2020.

\bibitem{Bian19}
Simeng Bian, Xi~Huang, and Ziyu Shao.
\newblock Online task scheduling for fog computing with multi-resource fairness.
\newblock In {\em IEEE 90th Vehicular Technology Conference (VTC2019-Fall)}, pages 1--5, 2019.

\bibitem{Kaur17}
Sawinder Kaur, Manojit Ghose, and Aryabartta Sahu.
\newblock Energy efficient scheduling of real-time tasks in cloud environment.
\newblock In {\em 2017 IEEE 19th International Conference on High Performance Computing and Communications; IEEE 15th International Conference on Smart City; IEEE 3rd International Conference on Data Science and Systems (HPCC/SmartCity/DSS)}, pages 178--185, 2017.

\bibitem{meta_heuristic_joint_task_scheduling}
Dabiah Alboaneen, Hugo Tianfield, Yan Zhang, and Bernardi Pranggono.
\newblock A metaheuristic method for joint task scheduling and virtual machine placement in cloud data centers.
\newblock {\em Future Generation Computer Systems}, 115:201–212, February 2021.

\bibitem{Lin22}
Weiwei Lin, Wentai Wu, and Ligang He.
\newblock An on-line virtual machine consolidation strategy for dual improvement in performance and energy conservation of server clusters in cloud data centers.
\newblock {\em IEEE Transactions on Services Computing}, 15(2):766--777, 2022.

\bibitem{power_container_ml}
Tarek Menouer, Otman Manad, Christophe C{\'e}rin, and Patrice Darmon.
\newblock Power efficiency containers scheduling approach based on machine learning technique for cloud computing environment.
\newblock In {\em Pervasive Systems, Algorithms and Networks}, pages 193--206. Springer, 2019.

\bibitem{Zhou19}
Liang Zhou, Laxmi~N. Bhuyan, and K.~K. Ramakrishnan.
\newblock Goldilocks: Adaptive resource provisioning in containerized data centers.
\newblock In {\em Proceedings of the IEEE International Conference on Distributed Computing Systems (ICDCS)}, pages 666--677, 2019.

\bibitem{Xu19}
M.~Xu and Rajkumar Buyya.
\newblock Brownoutcon: A software system based on brownout and containers for energy-efficient cloud computing.
\newblock {\em Journal of Systems and Software}, 155:91--103, 2019.

\bibitem{concurrent_container_heterogeneous}
Yang Hu, Huan Zhou, Cees de~Laat, and Zhiming Zhao.
\newblock Concurrent container scheduling on heterogeneous clusters with multi-resource constraints.
\newblock {\em Future Gener. Comput. Syst.}, 102(C):562–573, jan 2020.

\bibitem{Tan22}
Boxiong Tan, Hui Ma, Yi~Mei, and Mengjie Zhang.
\newblock A cooperative coevolution genetic programming hyper-heuristics approach for on-line resource allocation in container-based clouds.
\newblock {\em IEEE Transactions on Cloud Computing}, 10(3):1500--1514, 2022.

\bibitem{cubic_power_model}
Miyuru Dayarathna, Yonggang Wen, and Rui Fan.
\newblock Data center energy consumption modeling : A survey.
\newblock {\em IEEE Communications Surveys and Tutorials}, pages 1--1, January 2015.

\bibitem{Laabadi2018}
S.~Laabadi, M.~Naimi, H.~El Amri, and B.~Achchab.
\newblock The 0/1 multidimensional knapsack problem and its variants: A survey of practical models and heuristic approaches.
\newblock {\em American Journal of Operations Research}, 8:395--439, 2018.

\bibitem{piT}
Manojit Ghose, Sawinder Kaur, and Aryabartta Sahu.
\newblock Scheduling real time tasks in an energy-efficient way using vms with discrete compute capacities.
\newblock {\em Computing}, 102(1):263--294, 2020.

\bibitem{CloudSim}
Rodrigo~N. Calheiros, Rajiv Ranjan, Anton Beloglazov, César~AF De~Rose, and Rajkumar Buyya.
\newblock Cloudsim: A toolkit for modeling and simulation of cloud computing environments and evaluation of resource provisioning algorithms.
\newblock {\em Software: Practice and Experience}, 41(1):23--50, 2011.

\bibitem{clusterdata:Wilkes2011}
John Wilkes.
\newblock More {Google} cluster data.
\newblock Google research blog, November 2011.
\newblock Posted at \url{http://googleresearch.blogspot.com/2011/11/more-google-cluster-data.html}.

\end{thebibliography}
\bibliographystyle{unsrt}

\end{document}